\documentclass{sig-alternate}
\usepackage{stmaryrd}
\usepackage{latexsym}
\usepackage{epic}
\usepackage{eepic}
\usepackage{times}
\def\comment #1{}
\def\t{\hspace*{.5cm}}
\def\tt{\hspace*{1cm}}
\def\ttt{\hspace*{1.5cm}}
\def\tttt{\hspace*{2cm}}
\def\ttttt{\hspace*{2.5cm}}

\newtheorem{theorem}{Theorem}
\newtheorem{corollary}{Corollary}

\begin{document}
%

\title{Model-checking Driven Black-box Testing Algorithms
for  Systems with Unspecified Components}
\subtitle{[Extended Abstract]}

\numberofauthors{1}
\author{
       \alignauthor Gaoyan Xie {~and~} Zhe Dang\\     
       \affaddr{School of Electrical Engineering and Computer Science}\\
       \affaddr{Washington State University, Pullman, WA 99164, USA}\\
       \email{\{gxie,zdang\}@eecs.wsu.edu}
}
\maketitle

\begin{abstract}
Component-based software development has posed a serious challenge to
system verification since externally-obtained components could be a
new source of system failures. This issue can not be completely solved
by either model-checking or traditional software testing techniques
alone due to several reasons: (1) externally obtained components are
usually unspecified/partially specified; (2) it is generally difficult
to establish adequacy criteria for testing a component; (3) components
may be used to dynamically upgrade a system. This paper introduces a
new approach (called {\em model-checking driven black-box testing})
that combines model-checking with traditional black-box software
testing to tackle the problem in a complete, sound, and automatic
way. The idea is to, with respect to some requirement (expressed in
CTL or LTL) about the system, use model-checking techniques to derive
a condition (expressed in a communication/witness graph) for an
unspecified component such that the system satisfies the requirement
iff the condition is satisfied by the component. The condition's
satisfiability can be established by testing the component with
test-cases generated from the condition on-the-fly. In this paper, we
present algorithms for model-checking driven black-box testing, which
handle both CTL and LTL requirements for systems with unspecified
components. We also illustrate the ideas through some examples.
 
\end{abstract}




\pagenumbering{arabic}
\pagestyle{plain}

\section{Introduction}

\comment{
Model-checking  \cite{CE81,SC85,CES86,VW86,M93} 
has been proved to 
be an effective and powerful
 technique 
in verifying 
a finite-state system through exhaustive state exploration.
 However, there are still some fundamental
 issues that are hampering this technique from being a widely-accepted engineering practice
 in system development. One such issue arises from 
}

Component-based software development  \cite{KB98,BW98}
is a systematic engineering method to build software systems from prefabricated software
 components that are previously developed by the same organization,
 provided by third-party software vendors, or even purchased as
 commercial-off-the-shelf (COTS) products. Though this development
 method has gained great popularity in recent years, it has also posed serious challenges
 to the quality assurance issue of component-based software since
externally obtained
 components could be a new source of system failures. The issue is of vital importance to
 safety-critical and mission-critical systems. For instance, in June 1996,
 during the maiden voyage of the Ariane 5 launch
 vehicle, the launcher veered off course and exploded less than one minute after taking
 off. The report  \cite{Ariane} of the Inquiry Board indicates that the disaster
 resulted from insufficiently tested software reused from the Ariane 4. The developers
 had reused certain Ariane 4 software component in the Ariane 5 without substantially
 testing it in the new system, having assumed that there were no significant
 differences in these portions of the two systems.

Most of the current work addresses the issue from the viewpoint of
component developers:
how to
ensure the quality of components before they are released.
However, this view is obviously insufficient: an extensively
tested component (by the vendor) may still not perform as expected
in a specific deployment environment, since the systems where a
component could be deployed may be quite different and diverse and
they may not be tried out by its vendor. So, we look at this
issue from system developers' point of view:
\begin{quote}
(*) {\em how to ensure that a component functions correctly in the
host system where the component is deployed.}
\end{quote}
In practice, testing is almost the most natural resort to resolve
this
issue. When integrating a component into a system, system
developers may have three options for testing: (1) trust the
component provider's claim that the component has
 undergone thorough testing and then go ahead to use it;
(2) extensively retest the component alone; (3) hook the
component with the system and conduct integration testing.
Unfortunately, all of the three options have some serious
limitations. Obviously, for systems requiring high reliability,
the first option is totally out of the question. The second
option may suffer from the following fact. Software components
are generally built with multiple sets of functionality  \cite{GL02}, and
indiscriminately testing all the functionality of a software
component is not only expensive but sometimes also infeasible,
considering the potentially huge state space of the component interface.
 Additionally, it is usually difficult to know
when the testing over the component is adequate. The third option
is not always applicable. This is because, in many applications,
software components could be applied for  dynamic upgrading or
extending a running system  \cite{SZY03} that is costly or not
supposed to shut down for retesting at all. Even without all the
above limitations, purely testing-based strategies are still not
sufficient to establish the solid confidence for a reliable
component required by mission-critical or safety-critical systems,
where formal methods like model-checking are highly
desirable. However, one fundamental obstacle for using a formal
method to address the issue of (*) is that design details or
 source code of an externally obtained software component is generally
not fully available to the developers of its host system. Thus,
existing formal verification techniques (like model-checking) are
not directly applicable.

Clearly, this problem plagues both component-based software
systems and some hardware systems with a modularized design.
Generally, we call such systems as {\em systems with unspecified
components} (in fact, in most cases, the components are partially
specified to which our approach still applies.).

In this paper, we present 
 a new approach,
called {\em model-checking driven black-box testing},
 which combines model-checking techniques and black-box testing techniques 
to deal with this problem.
The idea is simple yet novel: with respect to some temporal requirement about
a system with an unspecified component, a model-checking based technique
is used to derive automatically a 
condition
 about the unspecified component
 from the rest of the system. This condition guarantees that 
the system satisfies the
requirement iff the condition is satisfied by the unspecified component, 
which can be checked
by adequate black-box testing over the unspecified component
 with test-cases generated automatically from the condition.

We provide algorithms for
both 
LTL and CTL model-checking driven black-box testing.
In the algorithms, the condition mentioned earlier
is represented as
communication graphs and witness graphs,
on which
a bounded and nested depth-first search procedure is employed to run
black-box testing over the unspecified component.
Our algorithms are both sound and complete.

Though we do not have an exact complexity analysis
result, our preliminary studies show that,
in the liveness testing algorithm for LTL,
the maximal length of test-cases run on the component is bounded
by $O(n\cdot m^2)$. For CTL, the length
is bounded by $O(k\cdot n\cdot m^2)$.
In here, $k$ is the number of CTL operators in the
formula to be verified, $n$ is the state number in the host system,
and $m$ is the state number in the component.

The advantages of our approach are
obvious: a stronger
confidence about the reliability of the system can be established
through both formal verification and adequate functional
testing; system developers can customize the testing with respect
to some specific system properties; intermediate model-checking
results (the communication and witness graphs) can be reused to avoid
(repetitive) integration testing when the component is updated, if
only the new component's interface remains the same;
 our algorithms are both sound and complete;
most of all, the whole process can be carried our in an
automatic way.

The rest of this paper is organized as follows. Section \ref{prel}
provides some background on temporal logics
LTL and CTL
along with our
model of 
systems containing unspecified components.
The main body of the paper 
consists of  Section \ref{ltltesting} and
Section \ref{ctltesting}, which 
propose algorithms for LTL and CTL model-checking driven black-box testing,
 respectively, 
over the system model.
Section \ref{examples}
illustrates
 the algorithms
through an example.
Section \ref{relatedwork}
lists some of the 
 related work.
Section \ref{discussions}
concludes the paper with 
some further issues to be resolved in the future.

Details on some algorithms are
omitted in this extended abstract. 
At http://www.eecs.wsu.edu/$\sim$gxie,
a full version of this paper is available.


\section{Preliminaries}\label{prel}

\subsection{The System Model}
In this paper, we consider systems with only one unspecified component
(the algorithms generalize to 
systems with multiple unspecified components). 
Such a system
is denoted 
by
 $$Sys=\langle M, X \rangle,$$ where $M$ is the host system and $X$ is the unspecified component.
Both $M$ and $X$ are finite-state transition systems communicating synchronously with each other via
 a finite set of input and output symbols.

Formally, the unspecified component
 $X$ is defined as a deterministic
 Mealy machine whose internal structure is unknown (but an implementation of $X$ is available for testing).
 We write $X$ as a triple $\langle\Sigma,
\nabla, m\rangle$, where $\Sigma$ is the set of $X$'s input symbols, $\nabla$ is the set of $X$'s
 output symbols, and $m$ is an upper bound for the number of states
in $X$
(as a convention in black-box testing,
the $m$ is given). 
Assume that $X$ has
 an initial state $s_{init}$.
A {\em run} of $X$ is
 a sequence of symbols alternately in $\Sigma$ and $\nabla$: $\alpha_0\beta_0\alpha_1\beta_1...$,
 such that, starting from the initial state $s_{init}$, $X$ outputs exactly the sequence $\beta_0\beta_1...$
  when it is given the sequence $\alpha_0\alpha_1...$ as input. In this case, we say that the input sequence is accepted by $X$.

The host system
$M$ is defined as a
$5$-tuple $$\langle S, \Gamma, R_{env}, R_{comm}, I\rangle$$ where
\begin{itemize}
\item  $S$ is a finite set of states;
\item  $\Gamma$ is a finite set of events;
\item  $R_{env}\subseteq S\times\Gamma\times S$ defines a set of {\em environment transitions} where $(s, a, s^\prime)\in R_{env}$
        means that $M$ moves from state $s$ to state $s^\prime$ 
upon receiving 
an event (symbol) $a\in\Gamma$ from the outside environment;
\item  $R_{comm}\subseteq S\times\Sigma\times\nabla\times S$ defines a set of {\em communication transitions} where
        $(s,\alpha,\beta,s^\prime)\in R_{comm}$ means that $M$ moves from state $s$ to state $s^\prime$ when
        $X$ outputs a symbol $\beta\in\nabla$ after $M$ 
sends
 $X$ an input symbol $\alpha\in\Sigma$; and,
\item  $I\subseteq S$ is $M$'s initial states.
\end{itemize}
Without loss of 
generality,
 we further assume that, there is only one transition between any two states
in $M$ (but $M$, in general, could still be  nondeterministic).

An {\em execution path} of the system $Sys=\langle M, X \rangle$ can be represented as a (potentially infinite) sequence
$\tau$  of states
 and symbols,
 $s_0c_0s_1c_1...$, where each $s_i\in S$, each $c_i$ is either a symbol in $\Gamma$
or a pair $\alpha_i\beta_i$ (called a {\em communication}) with $\alpha_i\in\Sigma$ and $\beta_i\in\nabla$. Additionally, $\tau$ satisfies
 the following requirements:
\begin{itemize}
\item $s_0$ is an initial state of $M$, i.e., $s_0\in I$;
\item for each $c_i\in\Gamma$, $(s_i,c_i,s_{i+1})$ is an environment transition of $M$;
\item for each $c_i=\alpha_i\beta_i$, $(s_i,\alpha_i,\beta_i,s_{i+1})$ is a communication transition of $M$.
\end{itemize}
\comment{let $\tau_{in}$ (resp. $\tau_{out}$) be the sequence obtained from $\tau$ by retaining
 only symbols in $\Sigma$ (resp. $\nabla$), then
 $\tau_{out}$ is the exact output of $X$ when it is given $\tau_{in}$ as input.}
The {\em communication trace} of $\tau$, denoted by $\tau_X$, is the sequence obtained from $\tau$
 by retaining only symbols in $\Sigma$ and $\nabla$
(i.e., the result of projecting $\tau$ onto $\Sigma$ and $\nabla$).
\comment{ An execution path is said to be {\em plain} if
its communication trace is empty.}
For any given state $s\in S$, we say that
the system $Sys$ can {\em reach} $s$
 iff $Sys$ has an execution path $\tau$ on which $s$ appears 
 and $\tau_X$ (if not empty) is also a run of $X$.

\comment{
such a computation on which $s$ appears. \comment{$\tau$: $s_0c_0s_1c_1....s$}
 An execution path is said to be {\em plain} if
its communication trace is empty.  An execution path is said to be a {\em computation} if it is
 plain, or its communication trace is a run of $X$.
}

\comment{(or, $s$ is {\em reachable} in $Sys$)}
\comment{For any given state $s\in S$, we say the system $Sys$ can {\em reach} $s$
 iff $Sys$ has such a computation on which $s$ appears. }\comment{$\tau$: $s_0c_0s_1c_1....s$.}
\comment{ We further say system $Sys$ can {\em obviously-reach}
 $s$ the computation\comment{(or, $s$ is {\em obviously-reachable} in $Sys$) iff $\tau$} is plain.}
In the case when $X$ is fully specified, the system can be regarded as
an I/O automaton \cite{lynch87hierarchical}.

\subsection{Model-checking}

Model-checking\cite{CE81,SC85,CES86,VW86,M93} is an automatic technique for verifying a finite-state system against some temporal 
specification. 
The system is usually represented by a Kripke structure $K=\langle S,R,L\rangle$ over a set of atomic
 propositions $AP$, where 
\begin{itemize}
\item $S$ is a finite set of states;
\item $R\subseteq S\times S$ is the (total) transition relation;
\item $L:S\rightarrow 2^{AP}$ is a function that labels each state with
 the set of atomic propositions that are  true in the
 state.
\end{itemize}

The temporal specification can  be expressed
in, among others,
 a branching-time temporal logic (CTL) or a linear-time temporal logic (LTL). 
\comment{
 in which there
 can be many possible futures at one time, or
in which there is only one future at one time. 
}
Both CTL and LTL formulas are composed of {\em path quantifiers} $A$ and $E$, which denote ``for all paths'' and ``there exists a
 path'',
 respectively, and {\em temporal operators} $X$, $F$, $U$ and $G$, which stands for ``next state'', ``eventually'', ``until'', and ``always'',
 respectively.

More specifically, CTL formulas are defined as follows:
\begin{itemize}
\item Constants $true$ and $false$, and every atomic proposition
in $AP$
 are CTL formulas;
\item If $f_1$ and $f_2$ are CTL formulas, then so are $\neg f_1$, $f_1\wedge f_2$, $f_1\vee f_2$, $f_1\rightarrow f_2$, $EX~f_1$, $AX~f_1$, $EF~f_1$, $AF~f_1$, $E[f_1~U~f_2]$, $A[f_1~U~f_2]$, $EG~f_1$, $AG~f_1$.
\end{itemize}
Due to duality,
any 
 CTL formula can be expressed in terms of 
$\neg,\vee, EX, EU$ and $EG$. A CTL model-checking problem, formulated as 
$$K,s\models f$$, is to check whether the CTL formula $f$ is true at a state $s$.
 For example, $AF~f$ is true at state $s$ if $f$ will be eventually true on all paths from $s$; $E[f~U~g]$ is true at state $s$ if there exists a path from $s$ on which $f$ is  true at each step until $g$ becomes true.  

LTL formulas,
on the other hand, are all in the form of $A~f$ where $f$ is a {\em path formula} defined as follows:
\begin{itemize}
\item Constants $true$ and $false$, and every atomic proposition
in $AP$
 are path formulas;
\item If $f_1$ and $f_2$ are path formulas, then so are $\neg f_1$, $f_1\wedge f_2$, $f_1\vee f_2$, $X~f_1$, $F~f_1$, $[f_1~U~f_2]$, $G~f_1$.
\end{itemize}
An LTL model-checking problem, formulated as $$K,s\models A~f$$, is to check whether the path formula $f$ is true on all paths from a state $s$.
 For example, $AFG~f$ is true at $s$ if on all paths from $s$, after a future point $f$ will be always true; $AGF~f$ is
 true at $s$ if on all paths from $s$, $f$ will be true infinitely often.

More detailed background in model-checking and temporal logics
can be found in the textbook \cite{CGP99}.
The system
$Sys=\langle M, X\rangle$ defined earlier
can be  understood as
a Kripke structure (with a given labeling function
and atomic propositions over states in $M$).
Since $X$ is an unspecified component,
in the rest of the paper, we mainly focus on 
how to solve the 
 LTL/CTL
model-checking problems
on the $Sys$ through black-box testing on $X$.

\subsection{Black-box Testing}

Black-box testing (also called functional testing) is a technique to test a system without knowing its internal structure.
 The system is regarded as a ``black-box'' in the sense that its behaviour 
can only be determined by
 \comment{It tries to study the system's behaviour only by} 
observing (i.e., testing)
its input/output sequences.
As a common assumption in
black-box testing, the unspecified component 
$X$ (treated as a black-box) has a special input
symbol $reset$
which
always makes $X$ return to its initial state
regardless of
its current state.
We use $Experiment(X,reset\pi)$
to denote the output sequence
obtained from the input sequence
$\pi$, when $X$ runs from the initial state (caused by the $reset$).
After running this $Experiment$, suppose that
we
continue to run $X$ by providing 
an input symbol $\alpha$ following the sequence $\pi$.
Corresponding
to this $\alpha$, we may obtain an output symbol $\beta$
from $X$.
We use $Experiment(X, \alpha)$ to denote the 
$\beta$.
Notice that this latter $Experiment$ is a shorthand for
``the last output symbol in \\ 
$Experiment(X, reset\pi\alpha)$".

Studies have shown that if only an upper bound for 
the number of states in the system
 and the system's inputs set are known, then its (equivalent) internal 
structure can be fully recovered
 through black-box testing.
Clearly, a naive algorithm
to solve the LTL/CTL model-checking problem
over the $Sys$ is to
first recover the full structure of the component $X$ through testing,
and then to
solve the classic model-checking problem over the fully specified
system composed from $M$ and the recovered $X$.
Notice that, in the naive algorithm,
when we perform black-box testing over $X$,
the selected test-cases have nothing to do with 
the host system $M$.
Therefore, it is desirable to find 
more sophisticated  algorithms
such as the ones discussed in this paper,
 that only select ``useful"
test-cases wrt the $M$ as well as the temporal specification of $M$ that 
needs to be checked.

\section{LTL Model-Checking Driven\\
 Black-Box Testing}
\label{ltltesting}

In this section, we introduce algorithms for LTL model-checking driven black-box testing
for the 
system $Sys=\langle M,X\rangle$ defined earlier.
 We first show how to solve a liveness analysis problem.
Then, we 
discuss the general LTL model-checking problem.

\subsection{Liveness Analysis}

The liveness analysis problem (also called 
the {\em infinite-often} problem)
 is to check: starting from some initial state $s_0\in I$, whether the system $Sys$ can
 reach a given state $s_f$ for infinitely many times.

 When $M$ has no communications with the unspecified
 component $X$, solving the problem is equivalent to 
  finding a path $p$ that runs from $s_0$ to $s_f$ and a loop $C$ that passes $s_f$. 
However, as far as communications are involved, the problem gets more complicated.
 The existence of the path $p$
 does not ensure that the system can indeed reach $s_f$ from $s_0$ (e.g., communications with $X$
 may never allow the system to take the necessary transitions to reach $s_f$). 
Moreover, the
 existence of the loop $C$ does not guarantee that the system can run along $C$ forever either
 (e.g., after running along $C$
 for three rounds, the system may be forced to leave $C$ by the communications with $X$).

\comment{
\begin{figure*}[tbhp]
\center
\input{newfig001.eepic}
\caption{Liveness analysis}
\end{figure*}
}

We approach this infinite-often problem in three steps. 
First,
 we look at whether a definite answer to the problem is possible.
If we can find a path from $s_0$ to $s_f$ and a loop from $s_f$ to $s_f$
 that involve only environment transitions, then the original problem
(i.e., the infinite-often problem)
 is definitely true.
If such a path and a
loop, no matter what transitions they may involve, do not exist at all, 
then the original problem is definitely false. 
If no definite answer is possible, we construct a directed graph
$G$ and use it to generate test-cases for the unspecified component
$X$.
The graph $G$, called a {\em communication  graph}, 
is a subgraph of $M$,
 represents all
 paths and loops in $M$ that
 could witness the truth of the problem (i.e., paths that run from $s_0$ to $s_f$ and loops that pass $s_f$).
The graph $G$ is defined as 
a pair $\langle
N,E\rangle$, where $N$ is a set of nodes
and  $E$ is a set of
edges.
 Each edge of $G$ is annotated either by a pair $\alpha\beta$
 that denotes a communication of $M$ with $X$, or has no annotation.
We construct $G$ as follows.
\begin{itemize}
\item Add one node to $G$ for each state in $M$ that is involved in some path between $s_0$
      and $s_f$ or in a loop that passes $s_f$;
\item Add one edge between two nodes in $N$ if $M$ has a 
transition between two states corresponding
 to the two nodes respectively. If the transition involves a
 communication with $X$, then annotate the edge with the communication symbols.
\end{itemize}
 It is easy to see that the liveness analysis problem is true if and only if the truth
 is witnessed by a path in $G$.
\comment{the system 
can reach $G$ has
  a path that begins from its initial node and passes its final node for 
infinitely many times,
 and the sequence of annotations collected from the edges along the path is a run of $X$.
}
Therefore, the last step is to check whether $G$ has a path along which the system can
 reach $s_f$ from $s_0$ first and then reach $s_f$
for infinitely many times. More details of this step
are  addressed in the next subsection. 

See appendix \ref{algcheckio} for details on the above operations.

\subsection{Liveness Testing}\label{TestLTL}

 To check whether the constructed 
communication graph
$G$ has a path that witnesses the truth of the original problem,
  the straightforward way is to try out all paths in $G$ and then check, whether along some path, the system can
 reach $s_f$ from $s_0$ first and then reach $s_f$
for infinitely many times. The check is done by testing $X$ with
 the communication trace of the path to see whether it is a run of $X$.
However, one difficulty is that $G$ may contain loops, and certainly we can only test
 $X$ with a finite communication trace. 
\comment{Fortunately, from a result in  \cite{PVY99}}Fortunately, the following observations are straightforward: 
\begin{itemize}
\item To check whether the system can reach $s_f$ from $s_0$, we only need to consider paths
 with length less than $mn_1$ where $n_1$ is the maximal number of communications on all {\em simple paths}
 (i.e., no loops on the path)
between $s_0$ and $s_f$ in $G$, and $m$ is an upper bound for the
number of states
in the unspecified component $X$; 
\item To check whether the system
 can reach from $s_f$ to $s_f$ for infinitely many times, we only need to
 make sure that the system can reach $s_f$ for $m-1$ times, and between $s_f$
and $s_f$, the system goes through a path no longer than $n_2$ that is the maximal number
 of communications on all {\em simple loops} (i.e., no nested loops along the loop) in $G$
 that pass $s_f$.
\end{itemize}

 Let $n=max(n_1,n_2)$.
The following procedure $TestLiveness$ uses a bounded and nested
depth-first 
search to
 traverse the graph $G$
while testing $X$. It first tests whether the system can reach $s_f$ from $s_0$
 along a path with length less than $mn$, then it tests whether the system can further reach $s_f$
 to $s_f$ for $m-1$ more times. 
The algorithm maintains
 a sequence of input symbols
 that has been successfully accepted by $X$, an integer variable $level$ that records how many communications have been gone
 through without reaching $s_f$, and an integer variable $count$ that indicates how many times $s_f$ has been reached.
 At each step, it chooses one candidate from
 the set of all possible input symbols at a node, and feeds the input sequence concatenated with the candidate input symbol to $X$.
 If 
the candidate input symbol and the 
 output symbol (corresponding to the candidate input symbol)
 of $X$  match
 the annotation of an edge originating from the node, 
the procedure moves forward to try the destination node of the edge with $level$ increased by 1. If there is no match, then the procedure tries other candidates.
 But before trying any other candidate, we need to bring $X$ to its initial state by sending it the special
 input symbol $reset$.
The procedure returns $false$ when all candidates are tried without a match, or when  
more than $mn$ communications have been gone through without reaching $s_f$.
After $s_f$ is reached, the procedure increases $count$ by $1$ and resets $level$ to $0$.
The procedure returns $true$ when it has already encountered $s_f$ for $m$ times.

\smallskip

{\bf Procedure} $TestLiveness(X,\pi,s_0,s_f,level,count)$

\t  {\bf If} $level>mn$ {\bf Then}

\tt      {\bf Return} $false$;

\t  {\bf Else If} $s_0=s_f$ {\bf Then}

\tt      {\bf If} $count >= m$ {\bf Then}

\ttt          {\bf Return} $true$;  

\tt      {\bf Else}

\ttt           $count := count+1$; $level := 0$;



\t  {\bf For each} $(s_0,s^\prime)\in E$ {\bf Do}

\tt      $Experiment(X, reset\pi)$;

\tt      {\bf If} $TestLiveness(X,\pi, s^\prime,s_f,level,count)$ {\bf Then}

\ttt          {\bf Return} $true$;
 


\t  $Inputs := \{\alpha|(s_0,\alpha\beta, s^\prime)\in E\}$;

\t  {\bf For each} $\alpha\in Inputs$ {\bf Do}

\tt     $Experiment(X, reset\pi)$;

\tt     $\beta := Experiment(X,\alpha)$;

\tt     {\bf If} $\exists s^\prime:(s_0,\alpha\beta, s^\prime)\in E$ {\bf Then}

\ttt         {\bf If} $TestLiveness(X,\pi\alpha,s^\prime,s_f,level+1,count)$

\tttt             {\bf Then} {\bf Return} $true$;




\t  {\bf Return} $false$.


\smallskip

\noindent In summary, 
our liveness testing algorithm to solve the 
liveness analysis problem has two
steps: (1) build the communication graph $G$; (2) return
the truth of 
$$TestLiveness(X,reset,s_0,s_f,level=0,count=0).$$

\subsection{LTL Model-Checking Driven Testing}

Recall that the LTL model-checking problem is, for a
Kripke structure $K=\langle S,R,L\rangle$ with a state $s\in S$ and
 a path formula $f$, to determine if $K, s\models A~f$.
Notice that $K, s\models A~f$ if and only if $K,s\models\neg
E~\neg f$. Therefore it is sufficient to only consider formulas
in the form $E~f$. The standard LTL model-checking algorithm \cite{CGP99} first
 constructs a {\em tableau} $T$ for the path formula $f$.
 $T$ is also a Kripke structure and includes {\em every} path that
 satisfies $f$. Then the algorithm composes
$T$ with $K$ and obtains another Kripke structure $P$ which
includes exactly the set of paths that are in both $T$ and $K$.
Thus, a state in $K$ satisfies $E~f$ if and only if it is
 the start of a path (in the composition $P$) that satisfies $f$.

Define $sat(f)$ to be the set of states in $T$ that satisfy $f$ and use
the convention that $(s,s^\prime)\in sat(f)$ if and only if $s^\prime\in
sat(f)$. The LTL model-checking problem can be summarized by
the following theorem  \cite{CGP99}:

\begin{theorem}
$K,s\models E~f$ if and only if there is a state $s^\prime$ in $T$
such that $(s,s^\prime)\in sat(f)$ and $P, (s, s^\prime)\models
EG~true$ under fairness constraints $\{sat(\neg(g~U~h)\vee
h)~|~g~U~h~$ occurs in $f \}$.
\end{theorem}

Note that the standard LTL model-checking algorithm still 
applies to the system $Sys=\langle M, X\rangle$, although it contains an
 unspecified component X. To see this, the construction of the tableau
$T$ from $f$ and the
definition of {\em sat} are not affected by the unspecified component $X$.
 The composition of
$Sys$ and $T$ is a new system $Sys^\prime=\langle P, X\rangle$ where 
$P$ is the composition of $M$ and $T$. Then one can show

\begin{corollary}
$\langle M,X\rangle,s\models E~f$ if and only if
there is a state $s^\prime$ in $T$ such that $(s,s^\prime)\in sat(f)$
 and $\langle P, X\rangle, (s, s^\prime)\models EG~true$ under
fairness constraints $\{sat(\neg(g~U~h)\vee h)~|~g~U~h$ occurs in
$f \}$.
\end{corollary}

Obviously, checking whether there is a state $s^\prime$ in $T$
such that $(s, s^\prime)\in sat(f)$ is trivial. To check whether
$\langle P, X\rangle, (s, s^\prime)\models EG~true$ under the fairness
constraints is
equivalent to
 checking whether there is computation in $\langle P, X\rangle$ 
that
starts from $(s,s^\prime)$ and on which the fairness constraints are
 true infinitely often. 
One can show that this is equivalent to 
the liveness analysis problem we studied in the previous subsection,
and thus, the LTL model-checking problem can be solved by extending
our algorithms for the liveness analysis problem.
 Moreover, the algorithms are both complete and sound.
\comment{
to solve the LTL model-checking problem, 
an LTL model-checking driven testing algorithm can be 
obtained from 
the liveness testing algorithm (details are omitted here).
Moreover, the liveness testing algorithm
and the 
LTL model-checking driven testing algorithm
are
both complete and sound.
}
\section{CTL Model-Checking Driven \\
Black-Box Testing}
\label{ctltesting}

In this section, we introduce algorithms for CTL
model-checking driven black-box testing
 for the system $Sys=\langle M,X\rangle$.

\subsection{Ideas}\label{Ideas}

Recall that the CTL model-checking problem
is, for a  Kripke structure $K=(S, R, L)$, a state $s_0\in S$,
 and a CTL formula $f$, to check whether $K,s_0\models f$ holds. The
standard algorithm \cite{CGP99} for this problem operates by searching 
 the structure
 and, during the search,
labeling
 each state
$s$ with the set of subformulas of $f$ that are true at $s$.
Initially, labels of $s$ are just $L(s)$. Then,
 the algorithm goes
through a series of stages---during the $i$-th stage, subformulas
with the $(i-1)$-nested CTL operators are processed. When a
subformula is processed, it is added to the labels for
 each state
where the subformula is true. When 
all the stages 
are completed,
the algorithm returns $true$ when $s_0$ is labeled
 with $f$, or $false$ otherwise.

However, if a system is not completely specified,
 the standard
 algorithm does not work.
This is because,
 in the system
 $Sys=\langle M,X\rangle$, transitions of $M$ may depend on
 communications with the unspecified component $X$. 
In this section, we adapt the standard CTL model-checking
algorithm \cite{CGP99} to handle
 the system $Sys$ (i.e., to check
 whether $$\langle M,X\rangle, s_0\models f$$ holds where $s_0$ is an initial state in $M$ and $f$ is a CTL formula
over $M$). 

 The new algorithm
\comment{, called
CTL
model-checking driven testing algorithm,}
 follows
a structure similar to the standard
 one. It also goes through a series of
 stages to search  $M$'s state space and label each state during the search.
However,
during a 
 stage,  processing the subformulas is rather involved, since the truth
of a subformula $h$ at a state $s$ 
can not be simply decided (it may depend on communications).
Similar to the algorithm for the 
liveness analysis problem,
our ideas
here are
to construct
a graph  representing
 all the paths that  witness the truth of $h$ at $s$.
But, the new algorithm  
is far more complicated than the 
liveness testing algorithm for LTL, since the truth of a CTL formula is 
usually 
witnessed by a tree instead of a single path.
In the new algorithm, processing each subformula $h$ is sketched as follows.

When $h$ takes the form of 
$EX~g$, $E[g_1~U~g_2]$, or $EG~g$, 
we construct a graph that represents exactly all the 
paths that  witness the truth of $h$ at
 some state. We call such a graph the subformula's {\em witness graph} (WG),
written as
 $\llbracket h\rrbracket$. We also call
 $\llbracket h\rrbracket$ an {\em EX graph}, an {\em EU graph}, or an {\em EG graph} if $h$ takes the form of $EX~g$,
$E[g_1~U~g_2]$, or $EG~g$,
 respectively. 

 Let $k$ be the total number of CTL operators in $f$. 
In the algorithm, 
 we  construct $k$ 
WGs,
  and for each WG,
 we assign it
with
 a unique
 ID number that ranges 
between $2$ and $k+1$.
(The ID number 1 is reserved for constant $true$.)
 Let ${\cal I}$ be the mapping 
from the  WGs to
 their IDs;
i.e.,
 ${\cal I}(\llbracket h\rrbracket)$ 
denotes
 the ID number of $h$'s witness graph, and ${\cal I}^{-1}(i)$ 
denotes 
the witness graph with
 $i$ as its ID number, $1<i\le k+1$.
We label a state $s$ with 
ID number
$1$ if $h$ is true at $s$ and the truth does not depend on communications between $M$ and $X$.
Otherwise, we label $s$ with $2\le i\le k+1$ if $h$ could be true at $s$ and the truth would be witnessed only by some paths 
which
 start from $s$ 
in ${\cal I}^{-1}(i)$ and,
on which, 
communications are involved.

When $h$ takes the form of 
a Boolean combination of subformulas using $\neg$ and $\vee$,
the truth of $h$
 at state $s$
is also  a logic combination of the truths of 
the
 component subformulas at the same state. 
To this end,
we label the state with an {\em ID expression} $\psi$ defined
 as follows:
\begin{itemize}
\item $ID := 1~|~2~|~\ldots~|~k+1$;

\item $\psi := ID~|~\neg \psi~|~\psi\vee \psi$.
\end{itemize}
Let $\Psi$ denote the set of all ID expressions. For each subformula $h$, 
we construct a labeling (partial)
 function $L_h:S\rightarrow \Psi$ to record
 the ID expression labeled to each state during 
the processing of the subformula $h$, and the labeling function
 is returned when the subformula is processed.

The detailed procedure, called $ProcessCTL$,
 for processing subformulas will be given in 
Section \ref{ProcessCTL}.
 After all subformulas are processed,
a labeling function $L_f$ for the outer-most subformula 
(i.e., $f$ itself) is returned. 
The algorithm returns $true$ when $s$ is labeled with $1$
 by $L_f$. It returns $false$ when $s$ is not labeled at all.
In other cases,
a testing procedure over $X$
is applied to 
check whether the ID expression labeled
in $L_f(s)$
could be evaluated true.
The procedure, called $TestWG$,
will be given in Section \ref{TestWG}.
 In summary,
the  \comment{CTL model-checking driven black-box testing}
 algorithm (to solve the CTL model-checking problem $\langle M,X\rangle, s_0\models f$) is sketched
as follows:

\smallskip


{\bf Procedure} $CheckCTL(M,X,s_0,f)$

\t    $L_f := ProcessCTL(M,f)$

\t    {\bf If} $s_0$ is labeled by $L_f$ {\bf Then}

\tt        {\bf If} $L_f(s_0)=1$ {\bf Then}

\ttt            {\bf Return} $true$; 

\tt        {\bf Else}

\ttt            {\bf Return} $TestWG(X, reset, s_0, L_f(s_0))$;

\t    {\bf Else} (i.e., $s_0$ is not labeled at all)

\tt       {\bf Return} $false$.

\smallskip


\subsection{Processing a CTL formula}\label{ProcessCTL}

Processing a CTL formula $h$
is implemented through a recursive procedure $ProcessCTL$.
 Recall that any CTL formula can be expressed in terms of $\vee$, $\neg$, 
$EX$, $EU$, and $EG$. Thus, at each intermediate step
 of the procedure,
 depending on whether the formula $h$ is atomic or takes 
one of the following
 forms: $g_1\vee g_2$, $\neg g$, $EX~g$,
$E[g_1~U~g_2]$, or $EG~g$,
the procedure 
has only six cases to consider and when it finishes, 
a labeling function $L_h$ is returned for formula $h$.

\smallskip

{\bf Procedure} $ProcessCTL(M,h)$

\t    {\bf Case} 

\tt   $h$ is atomic: Let $L_h$ label every state with 1 

\tttt whenever
$h$ is true on the state;

\tt   $h=g_1 \vee g_2$:  

\tttt                   $L_{g_1} := ProcessCTL(M,g_1)$;

\tttt                   $L_{g_2} := ProcessCTL(M,g_2)$;

\tttt                   $L_h := HandleUnion(L_{g_1}, L_{g_2})$;

\tt   $h=\neg g$:      

\tttt                   $L_g := ProcessCTL(M,g)$;

\tttt                   $L_h := HandleNegation(M, L_g)$;

\tt   $h=EX~g$:       

\tttt                   $L_g := ProcessCTL(M,g)$;

\tttt                   $L_h := HandleEX(M, L_g)$;

\tt   $h=E~[g_1 ~U~ g_2]$: 

\tttt                   $L_{g_1} := ProcessCTL(M,g_1)$;

\tttt                   $L_{g_2} := ProcessCTL(M,g_2)$;

\tttt                   $L_h := HandleEU(M, L_{g_1}, L_{g_2})$;

\tt   $h=EG~g$:       

\tttt                   $L_g := ProcessCTL(M,g)$;

\tttt                   $L_h := HandleEG(M,L_g)$;

\t    {\bf Return} $L_h$.

\smallskip

\noindent In the above procedure,
when $h=g_1 \vee g_2$,
we first process $g_1$ and $g_2$ respectively by 
calling $ProcessCTL$, then
 construct a labeling function $L_h$ for $h$ by 
merging (i.e., $HandleUnion$, see Appendix \ref{alghandleunion} for details))
$g_1$ and $g_2$'s labeling functions
$L_{g_1}$ and
$L_{g_2}$
as follows:
\begin{itemize}
\item For each state $s$ that is in both
$L_{g_1}$'s domain and
$L_{g_2}$'s domain,
let $L_h$ label $s$ with $1$ if either
$L_{g_1}$ or $L_{g_2}$ 
 labels $s$ with $1$ and label $s$ with 
ID expression $L_{g_1}(s)\vee L_{g_2}(s)$ otherwise;
\item For each state $s$ that is 
in $L_{g_1}$'s domain
(resp. $L_{g_2}$'s domain)
but not in $L_{g_2}$'s domain
(resp. $L_{g_1}$'s domain),
let $L$ label $s$ with
$L_{g_1}(s)$ (resp. $L_{g_2}(s)$).
\end{itemize}

When $h=\neg g$,
we first process $g$ by calling $ProcessCTL$, then
 construct a labeling function $L_h$ for $h$ by
``negating" (i.e., $HandleNegation$, see Appendix \ref{alghandlenegation} for details))
$g$'s labeling function $L_g$
as follows:
\begin{itemize}
\item For every state $s$ that is not in the domain of $L_g$, 
let $L_h$ label $s$ with $1$;
\item For each state $s$ that is in the domain of $L_g$ 
but not labeled with $1$ by $L_g$, let $L_h$ label $s$ with 
ID expression $\neg L_g(s)$.
\end{itemize}

The remaining three cases (i.e., 
for $EX$, $EU$, and $EG$)
 in the above procedure
are 
more complicated
 and are handled
in the following three subsections
respectively.

\subsubsection{Handling  EX}

When $h=EX g$, 
$g$ 
is processed first by $ProcessCTL$.
Then,
the  procedure
 $HandleEX$ is called with
$g$'s labeling function $L_g$ to 
construct a labeling function $L_h$ and create a 
witness graph for $h$ (we assume that, whenever a 
witness graph is created, the current value of a global variable
$id$, which initially is 2, is assigned as the ID number of the graph,
 and $id$ is incremented by $1$ after it is assigned to the graph).

The labeling function
$L_h$  is constructed as follows.
For each state $s$ that has a successor $s^\prime$ in 
the domain of $L_g$,
 if $s$ can reach $s^\prime$ through an environment transition and $s^\prime$ is labeled with $1$ by $L_g$ then let $L_h$ also label $s$ with $1$,
 otherwise let $L_h$ label $s$ with the current value of the global 
variable $id$.

 The witness graph for $h=EX g$,
 called an $EX$ graph, 
is created as 
a triple:
$$
\llbracket h\rrbracket = \langle N, E, L_g\rangle,
$$
 where $N$ is a set of nodes and $E$ is a set of
 annotated edges. It is created as follows:
\begin{itemize}
\item Add one node to $N$ for each state that is in the domain of $L_g$.
\item Add one node to $N$ for each state that has a
successor in the domain of $L_g$.
\comment{labeled by $L$ with the current value of $id$.} 
\item Add one edge between two nodes in $N$ to $E$ when $M$ has a 
transition between two states corresponding
 to the two nodes respectively; if the transition involves a
 communication with $X$ then annotate the edge with the communication symbols.
\end{itemize}
When $HandleEX$ finishes, it increases the global variable $id$ by $1$
(since one new witness graph has been created).

See Appendix \ref{alghandleex} for details.

\subsubsection{Handling  EU}

The case when $h=E~[g_1~U~g_2]$ is more complicated. 
We first process $g_1$ and $g_2$
 respectively by calling $ProcessCTL$, then call 
procedure
 $HandleEU$ with
$g_1$ and $g_2$'s labeling functions 
$L_{g_1}$ and $L_{g_2}$ to construct a labeling function
 $L_h$ and create a witness graph for $h$.

We construct the labeling function 
$L_h$ recursively. First,
 let $L_h$ label each state $s$ in the domain of $L_{g_2}$ with $L_{g_2}(s)$.
 Then,
 for state $s$ that has a successor $s^\prime$ in the domain of $L_{h}$, 
if both $s$ and $s^\prime$
 is labeled with $1$ by $L_{g_1}$ and $L_{h}$ respectively and 
$s$ can reach $s^\prime$ through an environment transition then
let $L_h$ also label $s$ with $1$, otherwise let $L_h$ label 
$s$ with the current value of the global variable $id$. 
Notice that, in the second step, if a state $s$ can be 
labeled with both $1$ and the current value of $id$, let $L_h$ label $s$ 
with $1$. Thus,
 we can ensure that the constructed $L_h$ is indeed
 a function.

 The witness graph for $h$, called an $EU$ graph, 
 is created
as a $4$-tuple:
$$
\llbracket h \rrbracket := \langle N, E, L_{g_1}, L_{g_2}\rangle,
$$
 where $N$ is a set of nodes and $E$ is a set of
 edges. $N$ is constructed by adding one node for each state 
that is in the domain of $L_h$, while $E$ is constructed 
in the same way as that of $HandleEX$.
When $HandleEU$ finishes, it increases the global variable $id$ by $1$.

See Appendix \ref{alghandleeu} for details.

\subsubsection{Handling  EG}

To handle formula $h=EG g$, we first process $g$ 
by calling $ProcessCTL$, then call 
procedure $HandleEG$ with
$g$'s labeling function $L_g$ to construct a 
labeling function $L_h$ and create
a witness graph for $h$.

The labeling function $L_h$
 is constructed as follows. For each state $s$ that can reach a 
loop $C$ through a path $p$ such that
 every state (including $s$) on $p$ and $C$ is in the domain of $L_g$, if every state (including $s$) on $p$ and $C$ 
is labeled with $1$ by $L_g$ and no communications are involved on the path and  the loop, then let $L_h$ also label $s$ with $1$, 
otherwise let $L_h$ label $s$ with the current value of the global 
variable $id$. 

 The witness graph for $h$, called an $EG$ graph,
 is created
as a triple:
$$
\llbracket h\rrbracket := \langle N, E, L_g\rangle,
$$
 where $N$ is a set of nodes and $E$ is a set of
 annotated edges. The graph is constructed in a same way as that of $HandleEU$.
When $HandleEG$ finishes, it also increases the global variable $id$ by $1$.

See Appendix \ref{alghandleeg} for details.

\subsection{Testing a Witness Graph}\label{TestWG}

As mentioned in Section \ref{Ideas},
the procedure for CTL model-checking driven black-box testing,
$CheckCTL$, consists of two parts.
The first part, which was discussed in 
Section \ref{ProcessCTL}, includes
$ProcessCTL$ that processes CTL formulas and creates
witness graphs.
The second part is to evaluate the created witness graphs
through testing $X$.
We will elaborate on this second part in this section.

In processing 
the CTL formula $f$,
a witness graph is constructed for each 
CTL operator in $f$ and
a labeling function is constructed for each subformula of $f$.
As seen from the algorithm
$CheckCTL$ (at the end of 
Section \ref{Ideas}),
the algorithm
either gives a definite ``yes'' or ``no'' answer to the 
CTL model-checking 
problem, i.e., $\langle M,X\rangle, s_0\models f$, 
or it reduces the problem to checking whether the ID expression 
$\psi$
labeled to $s_0$ can be evaluated true 
at the state. The evaluation procedure is carried out
by 
the following recursive procedure $TestWG$,
after an input sequence $\pi$ has been accepted by
 the unspecified component $X$.

\bigskip
\bigskip

{\bf Procedure} $TestWG(X,\pi,s_0,\psi)$

\t   {\bf Case}

\tt       $\psi=\psi_1\vee \psi_2$: 

\ttt          {\bf If} $TestWG(X,\pi,s_0,\psi_1)$ {\bf Then}

\tttt              {\bf Return} $true$;

\ttt          {\bf Else}
 
\tttt              {\bf Return} $TestWG(X,\pi,s_0,\psi_2)$

\tt       $\psi=\neg \psi_1$:

\ttt          {\bf Return} $\neg TestWG(X,\pi,s_0, \psi_1)$ 

\tt       $\psi=1$: 

\ttt          {\bf Return} $true$;   

\tt       $\psi=i$ with $2\le i\le k+1$: 

\ttt         {\bf When } ${\cal I}^{-1}(i)$ is an $EX$ graph

\tttt         {\bf Return} $TestEX(X,\pi,s_0,{\cal I}^{-1}(i))$; 

\ttt         {\bf When } ${\cal I}^{-1}(i)$ is an $EU$ graph

\tttt         {\bf Return} $TestEU(X,\pi,s_0,{\cal I}^{-1}(i), level=0)$; 

\ttt         {\bf When } ${\cal I}^{-1}(i)$ is an $EG$ graph

\tttt         {\bf Return} $TestEG(X,\pi,s_0,{\cal I}^{-1}(i)).$

\smallskip

\noindent In $TestWG$,
the first three cases are straightforward, which are consistent with
the intended meaning of ID expressions.
The cases
$TestEX, TestEU, TestEG$
 for  evaluating $EX, EU$, $EG$ 
graphs are discussed in the following three subsections.

\subsubsection{$TestEX$}

The case for checking whether an $EX$ graph $G=\langle N,E, L_g\rangle$ 
can be evaluated true at a state $s_0$ is simple. We just test whether the 
system $M$
 can reach from $s_0$ to another state $s^\prime\in {\bf dom}(L_g)$ 
through
a transition in $G$ such that the ID expression $L_g(s^\prime)$ 
can be  evaluated true
 at $s^\prime$.

See Appendix \ref{algtestex} for details.

\subsubsection{$TestEU$}
  
To check whether an $EU$ graph $G=\langle N,E, L_{g_1},
L_{g_2}\rangle$ can be evaluated true at a state $s_0$,
 we need to traverse all paths
$p$  in $G$ with length less than
 $mn$ and test the unspecified component $X$ to see 
whether the system can reach some state $s^\prime
\in {\bf dom}(L_{g_2})$
through
 one of those paths. 
In here,
$m$ is an upper bound for the number of states in the unspecified
 component $X$ and $n$ is the maximal number of communications on 
all simple paths between $s_0$ and $s^\prime$.
In the meantime, we should also check whether $L_{g_2}(s^\prime)$ 
can be evaluated true at $s^\prime$ and whether
 $L_{g_1}(s_i)$ can be evaluated true at 
$s_i$ for each $s_i$ on $p$ (excluding $s^\prime$) by calling $TestWG$.

See Appendix \ref{algtesteu} for details.
  
\subsubsection{$TestEG$}

For the case to check whether an $EG$ graph $G=\langle N,E, L_g\rangle$ 
can be evaluated true at a state $s_0$,
 we need to find an infinite path in $G$ along which the system can run forever.

 The following procedure $TestEG$ first
 decomposes $G$ into a set of SCCs.
Then, for each state $s_f$ in the SCCs, 
 it calls another procedure $SubTestEG$
to test whether the system can reach $s_f$
 from $s_0$ along a path not longer than $mn$, as well as 
whether the system can further reach $s_f$
 from $s_f$ for $m-1$ times. The basic idea of $SubTestEG$ (see
 Appendix \ref{algsubtesteg} for details) is 
similar to that of the $TestLiveness$ algorithm
 in Section \ref{TestLTL}, except that we need also check whether 
$L_g(s_i)$ can be evaluated true at $s_i$ for 
each state $s_i$ that has been reached so far by calling $TestWG$. 
Here, $m$ is the same as before while $n$ is the maximal number of
 communications on all simple paths between $s_0$ and $s_f$. 

\smallskip

{\bf Procedure} $TestEG(X,\pi,s_0,G=\langle N,E,L_g\rangle)$

\t   $SCC := \{C|C$ is a nontrivial SCC of $G\}$; 

\t   $T := \bigcup_{C\in SCC}\{s|s\in C\}$;

\t   {\bf For each} $s\in T$ {\bf Do} 

\tt       $Experiment(X,reset\pi)$;

\tt       {\bf If} $SubTestEG(X,\pi,s_0,s,G,level=0,count=0)$;

\ttt           {\bf Return} $true$;

\t   {\bf Return} $false$.

\smallskip

\noindent In summary, to solve the 
 CTL model-checking problem 
$$(M, X), s_0\models f,$$
  our algorithm 
$CheckCTL$ in Section \ref{Ideas}
either gives a definite yes/no
answer or
 gives a sufficient and necessary condition in the form of 
ID expressions
 and witness graphs.
 The condition
is evaluated through
black-box testing over the unspecified component
 $X$. The evaluation process 
will terminate with a yes/no answer to the model-checking problem.
One can show that 
our algorithm is 
both complete and sound.

\section{Examples}\label{examples}
In this section, to better understand our algorithms,
we look at some examples\footnote{The transition graphs in the figures in this section are not made total for the sake of readability.}.

\begin{figure}[tbhp]\comment{\label{figure01}}
\center
\includegraphics{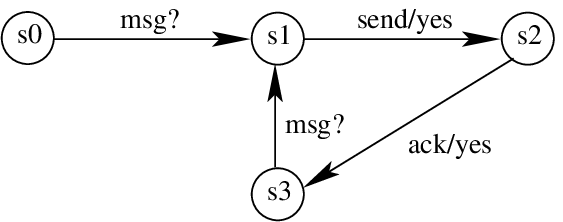}
\caption{}
\end{figure}

Consider a system $Sys=\langle M, X \rangle$ where $M$ keeps receiving messages from the outside environment
 and then transmits the message through the unspecified component $X$. The only event symbol in $M$ is $msg$,
 while $X$ has two input symbols $send$ and $ack$, and two output symbols $yes$ and $no$.
 The transition graph of $M$ is depicted in Figure 1\comment{\ref{figure01}} where we use a suffix $?$ to denote events 
 from the outside environment (e.g., msg?), and use a infix $/$ to denote communications of $M$
 with $X$ (e.g., $send/yes$). 

Assume that we want to solve the following LTL model-checking problem 
$$(M,X),s_0\models EGF s_2$$
 i.e., starting from the initial state $s_0$, the system can reach state $s_1$ infinitely often. Applying our liveness analysis algorithms, we can obtain the (minimized) communication graph in Figure 2.

\begin{figure}[tbhp]\comment{\label{figure01}}
\center
\includegraphics{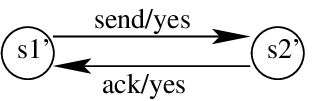}
\caption{}
\end{figure}

From this graph and our liveness testing algorithms, the system satisfies the liveness property iff the communication trace 
$$send~yes(send~yes~ack~yes)^{m-1}$$ 
is a run of $X$, where $m$ is an upper bound for number of states in $X$.
Now, we slightly modified the transition graph of $M$ into Figure 3 such that when a send fails, the system shall return to the initial state.

\begin{figure}[tbhp]\comment{\label{figure01}}
\center
\includegraphics{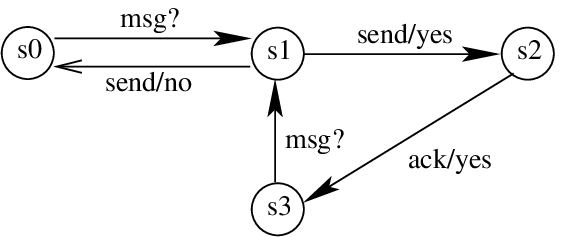}
\caption{}
\end{figure}

For this modified system, its (minimized) communication graph with respect to the liveness property would be as shown in Figure 4.
\begin{figure}[tbhp]\comment{\label{figure01}}
\center
\includegraphics{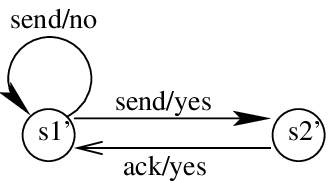}
\caption{}
\end{figure}

From Figure 4
 and the
 liveness testing algorithms, the system satisfies the liveness property iff there exist $0\le k_1, k_2\le 2m$ such that the communication trace 
$$(send~no)^{k_1} send~yes ((send~yes~ack~yes)(send~no)^{k_2})^{m-1}$$ is a run of $X$.

Still consider the system in Figure 3, but we want to solve a CTL model-checking problem $(M,X),s_0\models AF s_2$; i.e., along all paths from $s_0$, the system can reach state $s_1$ eventually. The problem is equivalent to 
$$(M,X),s_0\models \neg EG \neg s_2.$$
 Applying our CTL algorithms to formula $h=EG \neg s_2$, we construct an $EG$ witness graph $G=\langle N, E, L_{true}\rangle$ 
whose ID number is 
$2$ and a labeling function $L_h$, where $L_{true}$ labels all 
three states $s_0$,$s_1$, and $s_3$ with 
ID expression $1$
(as defined in Section \ref{Ideas}, which stands for $true$), 
and $L_{h}$ labels all three states $s_0$, $s_1$, and $s_3$ with $2$. The graph $G$ is depicted in Figure 5. From this graph as well as $L_h$, the algorithms conclude that 
 the model-checking
 problem is true iff the communication trace $(send~no)^{m-1}$ is not a run of $X$.
 
\begin{figure}[tbhp]\comment{\label{figure01}}
\center
\includegraphics{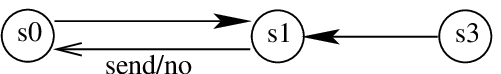}
\caption{}
\end{figure}

\begin{figure}[tbhp]
\center
\includegraphics{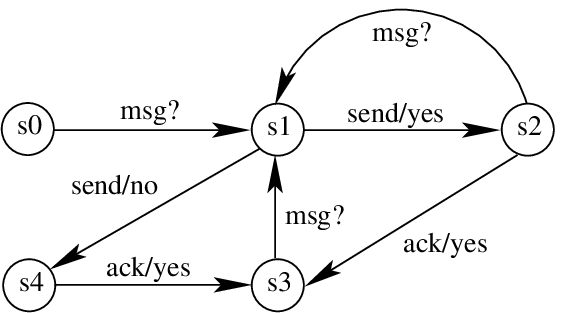}
\caption{}
\end{figure}

Now we modify the system in Figure 1
into a more complicated one
shown 
in Figure 6. 
For this system, we want to check 
$$(M,X),s_0\models \neg E [\neg s_2 U s_3]$$
i.e., starting from the initial state $s_0$, the system should never reach state $s_3$ earlier than it reaches $s_2$.
Applying our CTL algorithms to formula $$h=E[\neg s_2 U s_3],$$ 
we obtain 
an $EU$ witness graph $G=\langle N, E, L_1, L_2\rangle$ 
whose
ID number is 
 $2$ and a labeling function $L_h$, where $L_1$ labels all four states $s_0$, $s_1$, $s_3$ and $s_4$ with $1$, $L_2$ just labels $s_3$ with $1$, and $L_{h}$ labels states $s_0$, $s_1$, and $s_4$ with $2$, and labels $s_3$ with $1$. The graph $G$ is depicted in Figure 7. From this graph as well as $L_h$, the algorithms conclude that the model-checking problem is true iff none of communication traces in the form of $send~no(ack~yes~send~no)^*$ and with length less than $3m$ is a run of $X$.

\begin{figure}[tbhp]
\center
\includegraphics{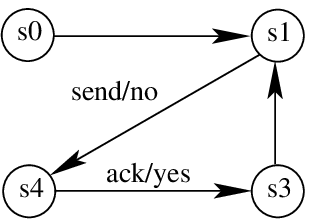}
\caption{}
\end{figure}

For the same system, we could 
consider more complicated
temporal 
properties as follows:

\begin{itemize}
\item $(M,X)\models AG (s_2\rightarrow AF s_3)$; 
i.e., starting from the initial state $s_0$, whenever the system reaches $s_2$, it would eventually reach $s_3$.
\item $(M,X),s_0\models AG (s_2\rightarrow AXA[\neg s_2 U s_3])$; i.e., starting from the initial state $s_0$, whenever it reaches state $s_2$, the system should never reach $s_2$ again until it reaches $s_3$.
\end{itemize}
We do
not include the witness graphs and labeling functions for these two cases 
in this extended abstract.
Nevertheless,
it can be concluded that
the two problems are true iff no communication traces with two consecutive symbol pairs $(send~yes)$ can be runs of $X$.

See Appendix \ref{example1} and Appendix \ref{example2} for details
about the above two examples.

\section{Related Work}\label{relatedwork}

The quality assurance problem for component-based software has
attracted lots of attention in the software engineering community,
as witnessed by recent publications in conferences like ICSE and FSE.
However, most of the 
work is
 based on the traditional testing techniques
and  considers
 the problem from the viewpoint
 of component developers;
i.e., how to ensure the quality of components before they are released.

Voas \cite{Voas98,Voas00}
 proposed a component certification strategy with the
 establishment of independent certification laboratories performing extensive testing
 of components and then publishing the results. Technically, this approach 
would not
 provide much improvement for solving the problem, since independent certification
 laboratories can not ensure the sufficiency of their testing either, and a testing-based
 technique alone is not enough to a reliable software component.
Some researchers
 \cite{SW00,OHR01} suggested an
 approach to augment a component with additional
 information to increase the customer's 
understanding and analyzing capability of the component behavior.
 A related approach \cite{WML02} is to automatically extract a 
finite-state machine model from 
 the interface of a software
 component, which  is delivered along with the component.
 This approach can provide some convenience for customers to test the component, but again,
 how much a customer should test is still a big problem.
To address the issue of testing adequacy, Rosenblum defined in \cite{Rosenblum97}
 a conceptual basis for testing component-based software, by introducing two notions
of $C$-{\em adequate}-{\em for}-${\cal P}$ and $C$-{\em adequate}-{\em for}-$M$ (with
respect to certain adequacy criteria) for adequate unit testing of a component and adequate
 integration testing for a component-based system,
 respectively. But this is still a purely testing-based strategy. In practice, how to
 establish the adequacy criteria is an unclear issue.

Recently, Bertolino et. al. \cite{BP03} recognized the importance of testing a software component
 in its deployment environment. They developed a framework that supports functional testing
of a software component with respect to customer's specification, which also provides a simple
 way to enclose with a component the developer's test suites which can be re-executed by the customer.
 Yet their approach requires the customer to have a complete specification about the component to be
 incorporated into a system, which is not always possible. McCamant and Ernst \cite{ME03} considered
 the issue of predicting the safety of dynamic component upgrade, which is part of the problem we
 consider. But their approach is completely different since they try to generate some abstract
 operational expectation about the new component through observing a system's run-time behavior
 with the old component.

In the formal verification area, there has been a long history of research on verification of systems
with modular structure (called modular verification \cite{Pnu85}).
 A key idea \cite{Lam83,henzinger98you} in modular verification is 
the {\em assume-guarantee} paradigm: A module should
 guarantee to have the desired behavior once the environment with
 which the module is interacting has the assumed behavior.
 There have been a variety of implementations for this idea
(see, e.g., \cite{AH+98}). 
However,
 the assume-guarantee idea does
not immediately
fit with our problem setup since it requires that users must have clear
 assumptions
 about a module's environment.

In the past decade, there has also been some research 
on combining model-checking and testing techniques
 for system verification, which can be classified into 
a broader class of techniques called specification-based testing. 
But most of the work only
 utilizes model-checkers' ability of
 generating counter-examples from a system's specification to produce test
 cases against an implementation \cite{CSE96,H97,EFM97,GH99,ABM98,BOY00,AB02}.

\comment{
Callahan et. al. \cite{CSE96} used the model-checker SPIN
\cite{H97} to check a program's execution
traces generated during white-box testing and to generate new test-cases
from the counter-example found by SPIN; in  \cite{EFM97}, SPIN was also
used to generate test-cases from counter-examples found during
model-checking system specifications. Gargantini and Heitmeyer \cite{GH99}
used SMV to both generate test-cases from the operational SCR
specifications and as test oracles. In \cite{ABM98,BOY00}, Ammann et. al.
 also exploited
the ability of
producing counter-examples with 
the model-checker  SMV \cite{M93};
but their approach is by mutating
both specifications and properties such that a large set of test
cases can be generated.
(A detailed introduction on using model-checkers
 in testing can be found in \cite{AB02}).
}

Peled et. al. \cite{PVY99,GPY02,Peled03CAV} studied the issue of checking a black-box against
 a temporal property (called black-box checking). But their focus is on how to
 efficiently establish an abstract model of the black-box through black-box testing
, and their approach requires a clearly-defined property (LTL formula) about the
black-box, which is not always possible in component-based systems.
Kupferman and Vardi \cite{KV97}
investigated module checking by considering the problem of checking 
an open finite-state system under {\em all} possible environments.
Module checking is different from 
the problem in (*)
 mentioned at the beginning of the paper
in the sense that
a component understood as an environment in \cite{KV97}
is a specific one.
Fisler et. al. 
\cite{FK01,LKF02} proposed
 an idea of deducing a model-checking condition for
 extension features from the base feature, which is adopted to study model-checking
 feature-oriented software designs. Their approach relies totally on
 model-checking techniques; their algorithms have false negatives and
 do not 
handle LTL formulas.

\section{Discussions}\label{discussions}

In this paper, we present algorithms for LTL and CTL model-checking driven black-box testing. The algorithms create communication graphs and witness graphs,
on which 
a bounded  and nested depth-first search procedure is employed to run 
black-box testing over the unspecified component.
Our algorithms are both sound and complete.
Though we do not have an exact complexity analysis
result, our preliminary studies show that,
in the liveness testing algorithm for LTL,
the maximal length of test-cases fed into the unspecified component
$X$ is bounded
by $O(n\cdot m^2)$. For CTL, the length
is bounded by $O(k\cdot n\cdot m^2)$.
In here, $k$ is the number of CTL operators in the
formula to be verified, $n$ is the state number in the host system,
and $m$ is the state number in the component.

The next natural step is to implement the algorithms
and see how well they  work in practice. 
In the implementation, there are further issues to be addressed.

\subsection{Practical Efficiency}

Similar to the traditional black-box testing 
algorithms to check conformance between Mealy machines,
the theoretical (worst-case)
 complexities are high in order to achieve
complete coverage.
However, worst-cases do not always occur in  a practical 
system. In particular, we need to identify scenarios
that our algorithms can be made more efficient.
For instance, using existing ideas of abstraction
\cite{cousotcousot-77}, we might obtain
a smaller but equivalent  
model of the host system before running the algorithms.
We might also, using additional partial information 
about the component, 
to derive a smaller state number for the component and to find ways to
expedite 
the model-checking process.
Notice that the number is actually
the state number for a minimal automaton that has the same set
input/output
sequences as the component.
Additionally, 
in the implementation, we also need a database to 
record the test results that have been performed so far
(so repeated testing can be avoided). 
Algorithms are needed to make use of the
test results to 
aggressively
trim the communication/witness graphs 
such that less test-cases are performed but the complete coverage is still
achieved. Also, we will study algorithms to minimize communication/witness graphs such that duplicate test-cases are avoided.
Lastly, it is also desirable to modify our algorithms such that
the communication/witness graphs
are generated  with the process of generating test-cases
and performing black-box testing over the unspecified component 
$X$.
In this way, a dynamic algorithm could be designed 
to trim the graphs
on-the-fly.

\comment{
\subsection{Symbolic Algorithms}
Recall that our problem scope is limited to finite-state systems,
and our algorithms
use standard finite set and relation operations. So they could 
be 
implemented in a symbolic way using techniques like BDDs \cite{B86},
combined with test-case generation.
It is is also interesting to see how 
the test-case generation and the
representation of the partial structure of the unspecified
component in our algorithms
are implemented in BDDs too.
}

\subsection{Coverage Metrics}

Sometimes, a complete coverage will not be achieved 
when running the algorithms on 
a specific application system.
In this case, a coverage metric
is needed to tell how much  
the test-cases that have run so far 
 cover. The metric will give a user
some confidence on the partial model-checking results.
Furthermore, such a metric would be useful in designing 
conservative algorithms to debug/verify the temporal specifications
that sacrifice the complete coverage but still 
bring the user
 reasonable confidence.

\subsection{More Complex System Models}

The algorithms can be generalized to systems containing 
multiple unspecified components.
Additionally, we will also consider cases when
these components interacts between each other, as well as cases
when the host system communicates with the components asynchronously.
Obviously, when the unspecified component (as well as  the host system)
 has
an infinite-state space, both the traditional model-checking
techniques and black-box techniques are not applicable. One
issue with infinite-state systems is that, the internal structure
of a general infinite-state system can not be learned through the
testing method.
Another issue is that model-checking a general
infinite-state system is an undecidable problem.
It is desirable to 
consider some restricted classes of 
infinite-state systems (such as real-time systems modeled
as timed automata \cite{AD94})
 where our algorithms 
generalize.
This is interesting, since through the study we may provide
an algorithm for model-checking driven black-box testing 
for a real-time system that contains an (untimed)
 unspecified component.
Since the algorithm will generate test-cases
for the component,
real-time integration testing over the composed
 system
 is avoided.

\bibliographystyle{abbrv}
\bibliography{fse04}

\appendix

\section{Definitions}

$R_{env}^s := \{(s,s^\prime)|\exists~a\in\Gamma:(s,a,s^\prime)\in R\}$;

$R_{comm}^s := \{(s,s^\prime)|\exists~\alpha\in\Sigma,\beta\in\nabla:(s,\alpha,\nabla,s^\prime)\in R_{comm}\}$;

$R^s := R_{env}^s\cup R_{comm}^s$;

$R_{env}^T := TranstiveClosure(R_{env}^s)$;
             
$R^T := TranstiveClosure(R^s)$;

{\bf Integer} $id := 1$;

\section{Algorithms}

\subsection{Liveness Analysis}\label{algcheckio}

\smallskip

{\bf Procedure} $CheckIO(\langle M, X\rangle, s_0, s_f)$

\t   $N := \emptyset$; $E:= \emptyset$;

\t   {\bf If} $(s_0,s_f)\in R_{env}^T\wedge(s_f,s_f)\in R_{env}^T$ {\bf Then}

\tt      {\bf Return} ``Yes'';

\t   {\bf Else if} $(s_0,s_f)\not\in R^T\wedge(s_f,s_f)\not\in R^T$ {\bf Then}

\tt      {\bf Return} ``No'';

\t   {\bf End if}

\t   $N:=\{s|(s_0, s)\in R^T\wedge(s,s_f)\in R^T)\}$;

\t   $E:=\{(s, s^\prime)|s,s^\prime\in N:(s,a,s^\prime)\in R_{env}\}\\\ttt
         \cup\{(s,\alpha\beta,s^\prime)|s,s^\prime\in N:(s,\alpha,\beta,s^\prime)\in R_{comm}\}$;

\t   {\bf Return} $TestIO(X,reset,s_0,s_f,level=0,count=0)$;

{\bf End procedure}

\subsection{Union of Labeling Functions}\label{alghandleunion}

\smallskip

{\bf Procedure} $Union(L_1,L_2)$

\t  $L:=\emptyset$;

\t  {\bf For each} $s\in {\bf dom}(L_1)\cup {\bf dom}(L_2)$ {\bf Do}

\tt      {\bf If} $s\in {\bf dom}(L_1)\cap {\bf dom}(L_2)$ {\bf Then}

\ttt          {\bf If} $L_1(s)=1\vee L_2(s)=1$ {\bf Then}

\tttt              $L := L\cup\{(s,1)\}$; 

\ttt          {\bf Else}

\tttt              $L := L\cup\{(s,L_1(s)\vee L_2(s))\}$;

\ttt          {\bf End if} 

\tt      {\bf Else if} $s\in {\bf dom}(L_1)$ {\bf Then} 

\ttt          $L := L\cup\{(s,L_1(s))\}$;

\tt      {\bf Else}

\ttt          $L := L\cup\{(s,L_2(s))\}$;

\tt      {\bf End if}

\t  {\bf End for}

\t  {\bf Return} $L$;

{\bf End procedure}

\subsection{Negation of a Labeling Function}\label{alghandlenegation}

\smallskip

{\bf Procedure} $Negation(M,L_1)$

\t  $L:=\emptyset$;

\t  {\bf For each} $s\in S$ {\bf Do}

\tt        {\bf If} $s\not\in {\bf dom}(L_1)$ {\bf Then}

\ttt            $L := L \cup\{(s,1)\}$;

\tt        {\bf Else if} $f(s)\not=1$ {\bf Then}

\ttt            $L := L \cup\{(s,\neg L_1(s))\}$;

\tt        {\bf End if}

\t  {\bf Return} $L$;

{\bf End procedure}

\subsection{Checking an EX Subformula}\label{alghandleex}

\smallskip

{\bf Procedure} $HandleEX(M,L_1)$

\t    $N:={\bf dom}(L_1)$; $L:=\emptyset$;

\t    {\bf For each} $t\in {\bf dom}(L_1)$ {\bf Do}

\tt         {\bf For each} $s:R^s(s,t)$ {\bf Do}

\ttt               $N := N\cup\{s\}$

\ttt               {\bf If} $L_1(t)=1\wedge R_{env}^s(s,t)$ {\bf Then}

\tttt                  {\bf If} $s\not\in {\bf dom}(L)$ {\bf Then}

\ttttt                          $L := L \cup \{(s, 1)\}$;

\tttt                   {\bf Else if} $L(s)\not=1$ {\bf Then}

\ttttt                          $L := L |_{s\leftarrow 1}$;

\tttt                   {\bf End if}

\ttt               {\bf Else if} $s\not\in{\bf dom}(L)$ {\bf Then}

\tttt                       $L := L\cup\{(s,id)\}$; 

\ttt               {\bf End for}

\tt         {\bf End for}

\t     {\bf End if}

\t    $E := \{(s,s^\prime)|s^\prime\in dom(f)\wedge\exists a:(s,a,s^\prime)\in R_{env}\}\\\ttt
            \cup\{(s,\alpha\beta,s^\prime)|s^\prime\in dom(f)\wedge(s,\alpha,\beta,s^\prime)\in R_{comm}\}$;

\t    {\bf Associate} $id$ {\bf with} $G=\langle N,E,L_1\rangle$; $id := id+1$;

\t    {\bf Return} $L$;

{\bf End procedure}

\subsection{Checking an EU Subformula}\label{alghandleeu}

\smallskip

{\bf Procedure} $HandleEU(M,L_1,L_2)$

\t  $L := L_2$;

\t  $T_1 := {\bf dom}(L_1)$; $T_2 := {\bf dom}(L)$;

\t  {\bf While} $T_2\not = \emptyset$  {\bf Do}

\tt       {\bf Choose} $t\in T_2$; $T_2 := T_2\setminus\{t\}$;

\tt       {\bf For each} $s\in T_1 \wedge R^s(s,t)$ {\bf Do}

\ttt            {\bf If} $L_1(s)=1\wedge L(t)=1\wedge R_{env}^s(s,t)$ {\bf Then}

\tttt                  {\bf If} $s\not\in {\bf dom}(L)$ {\bf Then}

\ttttt                          $T_2 := T_2 \cup \{s\}$; $L := L \cup \{(s, 1)\}$;

\tttt                   {\bf Else if} $L(s)\not=1$ {\bf Then}

\ttttt                          $T_2 := T_2 \cup \{s\}$; $L := L |_{s\leftarrow 1}$;

\tttt                   {\bf End if}

\ttt           {\bf Else if} $s\not\in {\bf dom}(L)$ {\bf Then}

\tttt               $T_2 := T_2 \cup \{s\}$; $L := L \cup \{(s, id)\}$;

\ttt          {\bf End if}

\tt      {\bf End for}

\t  {\bf End while}

\t  $N := {\bf dom}(L)$;

\t  $E := \{(s,s^\prime)|s,s^\prime\in N\wedge\exists a:(s,a,s^\prime)\in R_{env}\}\\\ttt
          \cup\{(s,\alpha\beta,s^\prime)|s,s^\prime\in N\wedge(s,\alpha,\beta,s^\prime)\in R_{comm}\}$;

\t  {\bf Associate} $id$ {\bf with} $G=\langle N,E,L_1,L_2\rangle$; $id := id+1$;

\t  {\bf Return} $L$;

{\bf End procedure}

\subsection{Checking an EG Subformula}\label{alghandleeg}

\smallskip

{\bf Procedure} $HandleEG(X,\pi,s_0,G=\langle N,E,L_g\rangle)$

\t   $SCC_{env} := \{C|C$ is a nontrivial SCC of $M$ and $C$ contains
no communication transitions $\}$; 

\t   $SCC_{comm} := \{C|C$ is a nontrivial SCC of $M$ and $C$ contains
some communication transitions $\}$; 

\t   $L := \{(s,1)|\exists C\in SCC_{env}:s\in C\}$\ttt

\t           $\cup\{(s,id)|\exists C\in SCC_{comm}:s\in C\}$

\t   $T := dom(L)$;

\t   {\bf While} $T\not=\emptyset$ {\bf Do}

\tt       {\bf Choose} $t\in T$; $T := T\setminus\{t\}$;

\tt       {\bf For each} $s\in {\bf dom}(L_1) \wedge R^s(s,t)$ {\bf Do}

\ttt              {\bf If} $L(t)=1\wedge L_1(s)=1\wedge R_{env}^s(s,t)$ {\bf Then}

\tttt                  {\bf If} $s\not\in {\bf dom}(L)$ {\bf Then}

\ttttt                          $T := T \cup \{s\}$; $L := L \cup \{(s, 1)\}$;

\tttt                   {\bf Else if} $L(s)\not=1$ {\bf Then}

\ttttt                          $T := T \cup \{s\}$; $L := L |_{s\leftarrow 1}$;

\tttt                   {\bf End if}

\ttt              {\bf Else if} $s\not\in {\bf dom}(L)$ {\bf Then}

\tttt                      $T := T \cup \{s\}$; $L := L \cup \{(s, id)\}$;

\ttt              {\bf End if}

\tt       {\bf End for}

\t   {\bf End While}

\t   $N := {\bf dom}(L)$;

\t   $E := \{(s,s^\prime)|s,s^\prime\in N\wedge\exists a:(s,a,s^\prime)\in R_{env}\}\\\ttt
           \cup\{(s,\alpha\beta,s^\prime)|s,s^\prime\in N\wedge(s,\alpha,\beta,s^\prime)\in R_{comm}\}$;

\t   {\bf Associate} $id$ {\bf with} $G=\langle N,E,L_1\rangle$; $id := id+1$;

\t   {\bf Return} $L$;

{\bf End procedure}

\subsection{Testing an EX Graph}\label{algtestex}

The algorithm for testing an $EX graph$ is simple. It first checks whether $L_1(s^\prime)$ can be evaluated true at
 any state $s^\prime$ such that the system can reach $s^\prime$ from $s_0$ through 
an environment transition. It returns true if it is the case. Otherwise,it chooses one candidate from
 the set of all possible input symbols from $s_0$, and feeds the
 sequence $\pi$ concatenated with the input symbol to $X$.
 If the output symbol of $X$ and the input symbol matches the annotation of an edge originating from the node, 
it moves forward to try the destination node of the edge. If there is no match,
 then it tries other candidates. But before trying any other
 candidate, it brings $X$ to its initial state by sending it the special input symbol,
 $reset$. The algorithm returns $false$ when all candidates are tried without a match.

\smallskip

{\bf Procedure} $TestEX(X,\pi,s_0,G=\langle N,E,L_1\rangle)$

\t  {\bf For each} $(s_0,s^\prime)\in E:s^\prime\in dom(L_1)$ {\bf Do}

\tt      $Experiment(X, reset\pi)$;

\tt      {\bf If} $TestWG(X,\pi, s^\prime, L_1(s^\prime))$ {\bf Then}

\ttt          {\bf Return} $true$;
 
\tt      {\bf End if}

\t  {\bf End for}

\t  $Inputs := \{\alpha|\exists \beta:(s_0,\alpha\beta,s^\prime)\in E\}$;

\t  {\bf For each} $\alpha\in Inputs$ {\bf Do}

\tt      $Experiment(X, reset\pi)$;

\tt      $\beta := Experiment(X, \alpha)$;

\tt     {\bf If} $\exists s^\prime:(s_0,\alpha\beta, s^\prime)\in E$ {\bf Then}

\ttt         {\bf If} $TestWG(X, \pi\alpha, s^\prime, L_1(s^\prime))$ {\bf Then}

\tttt             {\bf Return} $true$;

\ttt         {\bf End if}

\tt     {\bf End if}

\t  {\bf End for each};

\t  {\bf Return} $false$;

{\bf End procedure}

\subsection{Testing an EU Graph}\label{algtesteu}

The procedure $TestEU$ keeps a sequence of input symbols $\pi$
 that has been successfully accepted by $X$ and an integer $level$ that records how many communications have been gone
 through without reaching a destination state. And the algorithm works
 as follows.
At first, it checks whether it has gone through 
more than $mn$ communications without success, it returns false if it
 is the case. 
 Then, it checks whether it has reached a destination state (i.e.,
 $s_0\in dom(L_2)$). If it is the case,  it returns
 $true$ when $L_2(s_0)$ can be evaluated true $s_0$.
 Next, it checks whether $L_1(s_0)$ can be evaluated true at $s_0$, it
 returns false if it is not the case. 
 After that, it checks whether $L_1(s^\prime)$ can be evaluated true at
 any state $s^\prime$ such that the system can reach $s^\prime$ from $s_0$ through 
an environment transition. It returns true if it is the case. Otherwise,it chooses one candidate from
 the set of all possible input symbols from $s_0$, and feeds the
 sequence $\pi$ concatenated with the input symbol to $X$.
 If the output symbol of $X$ and the input symbol matches the annotation of an edge originating from the node, 
it moves forward to try the destination node of the edge with $level$ increased by 1. If there is no match,
 then it tries other candidates. But before trying any other
 candidate, it brings $X$ to its initial state by sending it the special input symbol,
 $reset$. The algorithm returns $false$ when all candidates are tried without a match.

\smallskip

{\bf Procedure} $TestEU(X, \pi, s_0, G=\langle
N,E,L_1,L_2\rangle,level)$

\t  {\bf If} $level>mn$ {\bf Then}\footnote{Here, $n$
  always denotes the  maximal number of communications on any simple paths in $G$.}

\tt      {\bf Return} $false$;

\t  {\bf Else if} $s_0\in dom(L_2)$ {\bf Then}

\tt      {\bf If} $TestWG(X,\pi, s_0, L_2(s_0))$ {\bf Then}

\ttt            {\bf Return} $true$;

\tt      {\bf End if}

\t  {\bf Else if not} $TestWG(X,\pi, s_0, L_1(s_0))$ {\bf Then}

\tt      {\bf Return} $false$;

\t  {\bf End if}
  
\t  {\bf For} $\exists s^\prime:(s_0,s^\prime)\in E$ {\bf Do}

\tt      $Experiment(X, reset\pi)$;

\tt      {\bf If} $TestEU(X,\pi,s^\prime,G,level)$ {\bf Then}

\ttt          {\bf Return} $true$;

\tt      {\bf End if}
 
\t  {\bf End for}

\t  $Inputs := \{\alpha|(s_0,\alpha\beta, s^\prime)\in E\}$;

\t  {\bf For each} $\alpha\in Inputs$ {\bf Do}

\tt     $Experiment(X, reset\pi)$;

\tt     $\beta := Experiment(X, \alpha)$;

\tt     {\bf If} $\exists s^\prime:(s_0,\alpha\beta, s^\prime)\in E$ {\bf Then}

\ttt         {\bf If} $TestEU(X,\pi\alpha,s^\prime,G,level+1)$ {\bf Then}

\tttt             {\bf Return} $true$;

\ttt         {\bf End if}

\tt     {\bf End if}

\t  {\bf End for each};

\t  {\bf Return} $false$;

{\bf End procedure}

\subsection{Subroutine for Testing an EG Graph}\label{algsubtesteg}

The procedure $SubTestEG$ keeps a sequence of input symbols
 that has been successfully accepted by $X$, an integer $level$ that records how many communications have been gone
 through without reaching $s_f$, and an integer $count$ that indicates how many times $s_f$ has been reached.
It first checks whether it has gone through 
more than $mn$ communications without reaching $s_f$, it returns false if it is the case. 
 Then, it checks whether it has reached the given state $s_f$. If it
 is the case, it
 returns $true$ when it has already reached $s_f$ for $m$ times, it
 increases $count$ by $1$ and resets $level$ to $0$ when otherwise.
 The next, it tests whether $L_1(s_0)$ can be evaluated true at $s_0$,
 and it returns false if it is not the case.
After that it checks whether $L_1(s^\prime)$ can be evaluated true at
 any state $s^\prime$ such that the system
 can reach $s^\prime$ from $s_0$ through 
an environment transition. It returns true if it is the
 case. Otherwise, it chooses one candidate from
 the set of all possible input symbols from $s_0$, and feeds the
 sequence $\pi$ concatenated with the input symbol to $X$.
 If the output symbol of $X$  and the input symbol matches the annotation of an edge originating from the node, 
it moves forward to try the destination node of the edge with level increased by 1. If there is no match,
 it tries other candidates. But before trying any other candidate, it brings
 $X$ to its initial state by sending it the special input symbol
 $reset$. The algorithm returns $false$ when all candidates are tried without a match.

\smallskip

{\bf Procedure} 

$SubTestEG(X,\pi,s_0,s_f,G=\langle N,E,L_1\rangle,level,count)$

\t  {\bf If} $level>mn$ {\bf Then} \footnotemark[\value{footnote}]

\tt      {\bf Return} $false$;

\t  {\bf Else if} $s_0=s_f$ {\bf Then}

\tt      {\bf If} $count >= m$ {\bf Then}

\ttt          {\bf Return} $true$;  

\tt      {\bf Else}

\tt           $count := count+1$; $level := 0$;

\tt      {\bf End if}

\t  {\bf Else if not} $TestWG(X,\pi, s_0, L_1(s_0))$ {\bf Then}

\tt      {\bf Return} $false$;

\t  {\bf End if}

\t  {\bf For} $\exists s^\prime:(s_0,s^\prime)\in E$ {\bf Do}

\tt      $Experiment(X, reset\pi)$;

\tt      {\bf If} $SubTestEG(X,\pi,s^\prime,s_f,G,level,count)$ {\bf Then}

\ttt          {\bf Return} $true$;

\tt      {\bf End if}
 
\t  {\bf End for}

\t  $Inputs := \{\alpha|(s_0,\alpha\beta, s^\prime)\in E\}$;

\t  {\bf For each} $\alpha\in Inputs$ {\bf Do}

\tt     $Experiment(X, reset\pi)$;

\tt     $\beta := Experiment(X,\alpha)$;

\tt     {\bf If} $\exists s^\prime:(s_0,\alpha\beta, s^\prime)\in E$ {\bf Then}

\ttt         {\bf If} $SubTestEG(X,\pi\alpha,s^\prime,s_f,G,level+1,count)$ {\bf Then}

\tttt             {\bf Return} $true$;

\ttt         {\bf End if}

\tt     {\bf End if}

\t  {\bf End for};

\t  {\bf Return} $false$;

{\bf End procedure}

\section{Examples}

\subsection{Check $(M,X)\models AG(s_2\rightarrow AF s_3)$} \label{example1}

To check whether $(M,X)\models AG(s_2\rightarrow AF s_3)$, is
equivalent to checking whether $$(M,X)\models \neg E[true~U(s_2\wedge
EG\neg s_3)].$$ We describe how the formula $$f=E[true~U(s_2\wedge EG\neg s_3)]$$ is
processed by $HandleCTL$ from bottom to up as
follows.

\begin{enumerate}
\item the atomic subformula $s_2$ is processed by $HandleCTL$, and a labeling
function $L_1=\{(s_2,1)\}$ is returned;
\item the atomic subformula $s_3$ is processed, and a labeling
function $L_2=\{(s_3,1)\}$ is returned;
\item to process $\neg s_3$, $HandleNegation$ is called with $L_2$ to return a labeling
function $L_3=\{(s_0,1),(s_1,1),(s_2,1),(s_4,1)\}$;
\item to process $EG\neg s_3$, $HandleEG$ is called with $L_3$ to construct an $EG~
graph$ $G_1=\langle N,E,L_3\rangle$ with id $2$ (see Figure 8) and
return a labeling function $L_4=\{(s_0,2),(s_1,2),(s_2,2)\}$;

\begin{figure}[tbh]
\center
\includegraphics{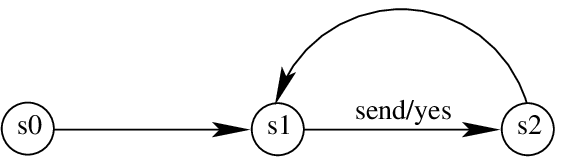}
\caption{}
\end{figure}

\item to process $s_2\wedge EG\neg s_3$, $HandleNegation$ and
  $HandleUnion$ are called with $L_1$ and $L_4$ to return a labeling function $L_5=\{(s_2,2)\}$;
\item to process $E[true~U(s_2\wedge EG\neg s_3)]$, $HandleEU$ is
  called with $L_5$ to construct an $EU graph$ $G_2=\langle N,E,L_5\rangle$ with id
  $3$ (see Figure 9) and return a labeling function $$L_f=\{(s_0,3),(s_1,3),(s_2,3),(s_3,3),(s_4,3)\}.$$

\begin{figure}[tbh]
\center
\includegraphics{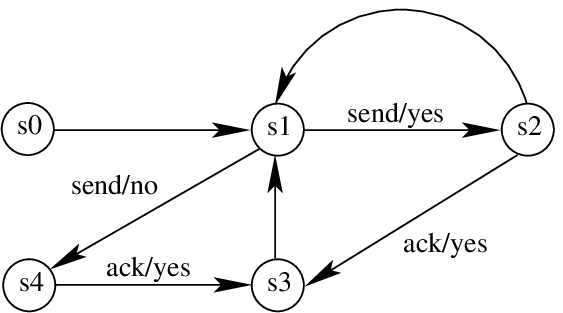}
\caption{}
\end{figure}

\end{enumerate}

Since $s_0$ is labeled by $L_f$ with an ID expression $3$ instead of $1$ (i.e.,
$true$), we need to test whether the ID expression $3$ can be
evaluated true at $s_0$ by calling $TestWG$ with $s_0$ and $G_2$. It's
easy to see that, essentially $TestWG$ would be testing whether some
communication trace (with bounded length) with two consecutive symbol pairs $(send~yes)$
is a run of $X$. It returns $false$ if such trace exists, or vice versa.

\subsection{Check $(M,X),s_0\models AG(s_2\rightarrow AXA[\neg s_2 U
  s_3])$} \label{example2}

To check whether $(M,X),s_0\models AG (s_2\rightarrow AXA[\neg s_2 U s_3])$, is
equivalent to checking whether 
$$(M,X)\models \neg E[true~U (s_2\wedge EX(E[\neg s_3 U (s_2\wedge\neg
s_3)]\vee EG\neg s_3))].$$ 
We describe how the formula 
$$f=E[true~U (s_2\wedge EX(E[\neg s_3 U (s_2\wedge\neg s_3)]\vee EG\neg s_3))]$$
 is processed by $HandleCTL$ from bottom to up as follows.

\begin{enumerate}
\item the atomic subformula $s_2$ is processed by $HandleCTL$, and a labeling
function $L_1=\{(s_2,1)\}$ is returned;
\item the atomic subformula $s_3$ is processed, and a labeling
function $L_2=\{(s_3,1)\}$ is returned;
\item to process $\neg s_3$, $HandleNegation$ is called with $L_2$ to return a labeling
function $L_3=\{(s_0,1),(s_1,1),(s_2,1),(s_4,1)\}$;

\item to process $s_2\wedge\neg s_3$, $HandleNegation$ and
  $HandleUnion$ are called with $L_1$ and $L_3$ to return a labeling
  function $L_4=\{(s_2,1)\}$;

\item to process $E[\neg s_3 U(s_2\wedge\neg s_3)]$, $HandleEU$ is
  called with $L_3$ and $L_4$ to construct an $EU~graph$ $G_1=\langle N,
  E, L_3, L_4\rangle$ with id $2$ (see Figure 10) and
return a labeling function $L_5=\{(s_0,2),(s_1,2),(s_2,1)\}$;

\begin{figure}[tbh]
\center
\includegraphics{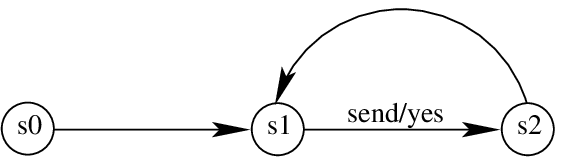}
\caption{}
\end{figure}

\item to process $EG\neg s_3$, $HandleEG$ is
  called with $L_3$ to construct an $EG~graph$ $G_2=\langle N,
  E, L_3\rangle$ with id $3$ (see Figure 11) and
return a labeling function $L_6=\{(s_0,3),(s_1,3),(s_2,3)\}$;

\begin{figure}[tbh]
\center
\includegraphics{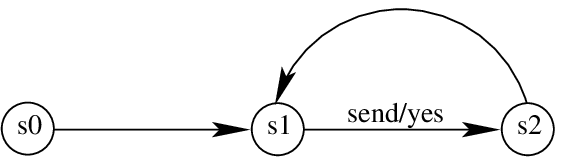}
\caption{}
\end{figure}

\item to process $E[\neg s_3 U(s_2\wedge\neg s_3)]\vee EG\neg s_3$,
  $HandleUnion$ is called with $L_5$ and $L_6$ to return a 
labeling function $L_7=\{(s_0,2\vee 3),(s_1,2\vee 3),(s_2,1)\}$;

\item to process $EX(E[\neg s_3 U(s_2\wedge\neg s_3)]\vee EG\neg s_3)$,
 $HandleEX$ is called with $L_7$ to construct an $EX~graph$ $G_3=\langle N,
  E, L_7\rangle$ with id $4$ (see Figure 12) and
return a labeling function $L_8=\{(s_0,4),(s_1,1),(s_2,4),(s_3,4)\}$;

\begin{figure}[tbh]
\center
\includegraphics{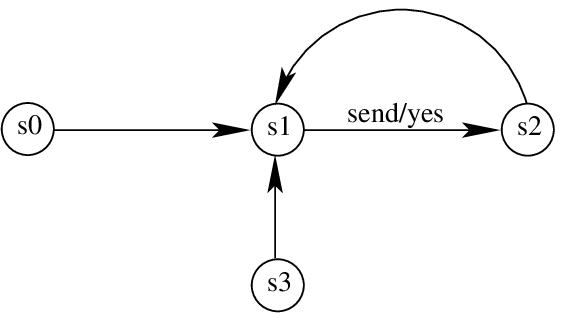}
\caption{}
\end{figure}

\item to process $s_2\wedge EX(E[\neg s_3 U(s_2\wedge\neg s_3)]\vee
  EG\neg s_3)$, $HandleNegation$ and
  $HandleUnion$ are called with $L_1$ and $L_8$ to return a labeling function $L_9=\{(s_2,4)\}$;

\item to process $E[true~U(s_2\wedge EX(E[\neg s_3 U(s_2\wedge\neg s_3)]\vee
  EG\neg s_3))]$, $HandleEU$ is
  called with $L_9$ to construct an $EU graph$ $G_4=\langle N,E,L_5\rangle$ with id
  $5$ (see Figure 13) and return a labeling function $$L_f=\{(s_0,5),(s_1,5),(s_2,5),(s_3,5),(s_4,5)\}.$$

\begin{figure}[tbh]
\center
\includegraphics{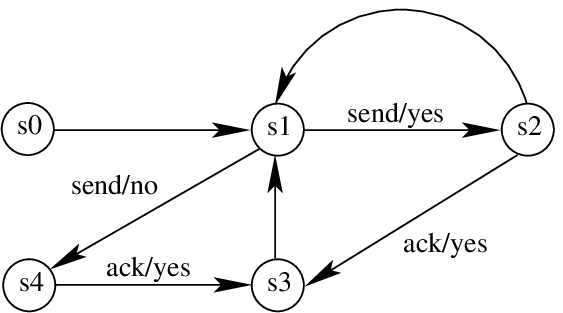}
\caption{}
\end{figure}

\end{enumerate}

Since $s_0$ is labeled by $L_f$ with an ID expression $5$ instead of $1$ (i.e.,
$true$), we need to test whether the ID expression $5$ can be
evaluated true at $s_0$ by calling $TestWG$ with $s_0$ and $G_4$. It's
easy to see that, essentially $TestWG$ would be testing whether some
communication trace (with bounded lengtg) with two consecutive symbol pairs $(send~yes)$
is a run of $X$. It returns $false$ if such trace exists, or vice versa.

\end{document}